\documentclass{appolb}
\usepackage{epsfig}
\usepackage{amssymb}

\newcommand{\muegamma}{\mu\to\e\gamma}
\newcommand{\mueee}{\mu\to\e\e\overline{\e}}
\newcommand{\mueN}{\mu{\rm N\to\e N}}
\begin{document}
\pagestyle{plain}
\newcount\eLiNe\eLiNe=\inputlineno\advance\eLiNe by -1
\title{Lepton Flavor Violation in Little Higgs Models
\thanks{Presented at XXXIII International Conference of Theoretical Physics (Ustro\'n 2009).  Talk also given a the Corfu Summer Institute (2009) by J.I. Illana.}
\author{Jos\'e Ignacio Illana, Mark D. Jenkins\footnote{Speaker.}}
\address{CAFPE and Departamento de F\'isica Te\'orica y del Cosmos, \\ Universidad de Granada,
E-18071 Granada, Spain}}
\maketitle

\begin{abstract}
We report on our study of the LFV processes $\muegamma$, $\mueee$ and $\mueN$ in the context of Little Higgs models.
Specifically we examine the \emph{Littlest} Higgs with T-parity (LHT) and the \emph{Simplest} Little Higgs (SLH) as examples of a Product group and Simple group Little Higgs models respectively.  The necessary Feynman rules for both models are obtained in the 't Hooft Feynman Gauge up to order $v^2/f^2$ and predictions for the branching ratios and conversion rates of the LFV processes are calculated to leading order (\emph{one-loop} level).  Comparison with current experimental constraints show that there is some tension and, in order to be within the limits, one requires a higher breaking scale $f$, alignment of the heavy and light lepton sectors or almost degenerate heavy lepton masses.  These constraints are more demanding in the SLH than in the LHT case.
\end{abstract}

\section{Introduction}

Little Higgs models \cite{ArkaniHamed:2001ca} introduce a new approximate global symmetry into the lagrangian, a subgroup of which is gauged.  This symmetry is spontaneously broken at a heavy scale $f$ of order 1 TeV.  The new gauge group breaks down to the SM gauge group while the SM Higgs doublet then appears as a pseudo-Goldstone boson of this global symmetry.  The low energy degrees of freedom are given by a non-linear sigma model which make this an effective theory valid below the cutoff $\Lambda = 4\pi f \sim 10$ TeV.  Beyond this cutoff one needs an ultraviolet completion.

The new global symmetry is explicitly broken by gauge and Yukawa interactions through a mechanism known as \emph{collective symmetry breaking}.  This allows us to give the Higgs a mass and non-derivative interactions but preserves the cancellation of the quadratic divergences at the one-loop level.  The dependence on the cutoff reappears at the two-loop level but this sensitivity to a 10 TeV cutoff is not \emph{unnatural}.

The same as in most new physics models, Little Higgs type models all introduce new gauge bosons and heavy fermions which, in this case, have masses of the order of the heavy breaking scale $f \sim 1$ TeV.  
These end up giving new contributions to lepton flavor violating processes which we have analyzed and compared with current experimental limits.  We compare two different models:  the \emph{Littlest} Higgs with T-parity (LHT) \cite{ArkaniHamed:2002qy} and the \emph{Simplest} Little Higgs \cite{Kaplan:2003uc} which are examples of Product group and Simple group models, respectively.

The LHT model introduces, among other particles, heavy partners to the standard model fermions as well as adding four new gauge bosons: $W_H^\pm$, $Z_H$ and $A_H$.  The new mixing in the lepton sector comes from the misalignment of the heavy and light Yukawa matrices.  This is analogous to the mechanism that produces the quark sector mixing (CKM matrix) in the Standard Model.

In the SLH case, each of the lepton families is enlarged to contain an additional heavy neutrino.  Again there are new heavy gauge bosons in this model ($X^\pm$, $Z'$, $Y^0$ and ${Y^0}^\dag$).  The mixing of the heavy neutrinos with the light neutrinos in conjunction with a family mixing in the light sector produces the new lepton mixing matrix.  

\begin{figure}[htp]
\centering
\begin{tabular}{rccl}
$\muegamma$: &
\raisebox{-11mm}{\includegraphics[scale=0.65]{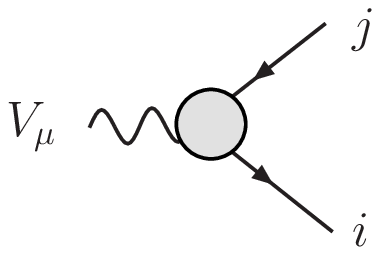}} &
\\ & {\em vertex} (triangles) & & \\
$\mueee$: &
\raisebox{-1cm}{\includegraphics[scale=0.65]{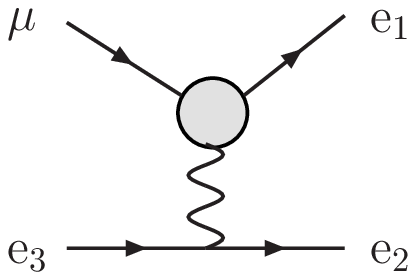}} &+
\raisebox{-1cm}{\includegraphics[scale=0.65]{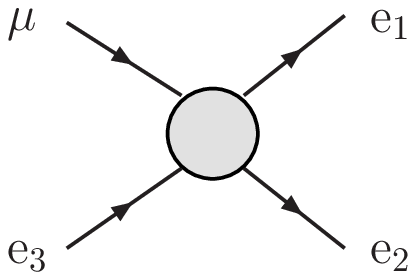}} &  
\hspace{-1cm}$+(\e_1\leftrightarrow\e_2)$
\\ & {\em $V$-penguins} (triangles+SE)& {\em e-boxes} \\
$\mueN$: &
\raisebox{-1cm}{\includegraphics[scale=0.65]{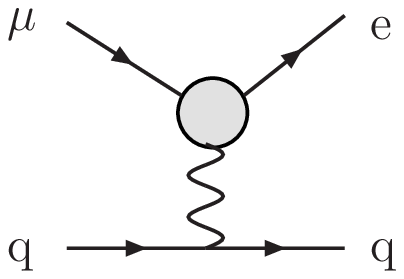}} &+
\raisebox{-1cm}{\includegraphics[scale=0.65]{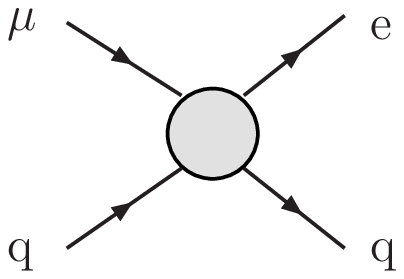}} & \hspace{-1cm}(q $=$ u, d)
\\ & {\em $V$-penguins} (triangles+SE)& {\em q-boxes} & 
\end{tabular}
\caption{Topologies contributing to LFV processes}
\label{topologies}
\end{figure}

\section{Lepton Flavor Violating processes and their contributions}
The leading order contributions to the LFV processes are all one-loop diagrams.  The topologies contributing to each of the three processes we are interested in are shown in figure \ref{topologies}.

The current experimental limits on these processes are $\mathcal{B}(\muegamma) < 1.2\times 10^{-11}$, \mbox{$\mathcal{B}(\mueee) < 10^{-12}$} and \mbox{$\mathcal{CR}(\mu{\rm Ti} \to {\rm eTi}) < 4.3\times 10^{-12}$} \cite{Amsler:2008zzb}.  The standard model predictions are much smaller than these limits (order $\sim 10^{-50}$).  This is because the effect is proportional to the neutrino masses divided by the $W^\pm$ mass to the fourth power.  This means that practically any LFV signal is a sure sign of physics beyond the Standard model.

\section{Results and conclusions}
Our procedure to obtain the LFV contributions in the two Little Higgs models was as follows \cite{delAguila:2008zu,newpaper}:
\begin{enumerate}
  \item We obtained the contributions of all triangle, self-energy and box diagrams in terms of generic couplings and standard loop integrals (Passarino-Veltman reduction).
  \item We worked with the Little Higgs lagrangians of both the LHT and SLH models in order to obtain the relevant Feynman rules for our processes in the 't Hooft-Feynman gauge.  This calculation is done up to order $v^2/f^2$ for all relevant couplings.  
  \item We substitute the couplings in our generic expresions and simplify as much as possible to obtain the form factors for each process \mbox{analytically}.
  \item Finally the form factors are introduced in to the expressions for the amplitudes and evaluated.
\end{enumerate}
In our calculation we find that all contributions are ultraviolet finite in both models for all three processes.

The Feynman rules for the LHT model had been obtained previously in \cite{Blanke:2006eb} and a first phenomenological study on the lepton sector was carried out in \cite{Blanke:2007db}.  The impact of a missing $v^2/f^2$ term pointed out in \cite{delAguila:2008zu} (and in \cite{Goto:2008fj} for the quark sector) was later assessed in \cite{Blanke:2009am} where the results of \cite{delAguila:2008zu} were confirmed.

To do our numerical analysis we made the simplification of using only two lepton generations.  This leaves us with four free parameters that we can vary.  The idea is to set three of the parameters to \emph{natural values} and move the fourth parameter.  The four parameters are:
\begin{itemize}
  \item The breaking scale $f$.
  \item The mixing angle $\theta$ appearing in the (now $2 \times 2$) mixing matrix $V^\ell_H$.
  \item An average mass of the two heavy leptons given by $\widetilde{y} = {m_{H1}m_{H2}} / M_G^2$ where $M_G$ is the mass of the charged heavy gauge boson ($X^\pm$ in SLH and $W_H^\pm$ in LHT) and $m_{Hi}$ is the mass of the heavy lepton.
  \item A relative splitting between the heavy lepton masses $\delta = \frac{m_{H2}^2-m_{H1}^2}{m_{H1}m_{H2}}$.
\end{itemize}
The natural values for these parameters are $f\sim 1$ TeV, $\theta \sim \pi/4$, $\widetilde{y} \sim 1$ and $\delta \sim 1$.
Other parameters have also been set to typical values.  Specifically, the heavy quark mases appearing in $\mueN$ are all set to $m_Q = 500$ GeV and we have assumed minimal mixing in the quark sector.  In the SLH model there is an extra parameter $\beta$ that has been set to $\tan \beta = 1$.

\begin{figure}[!htp]
\centering
\includegraphics[scale=0.71]{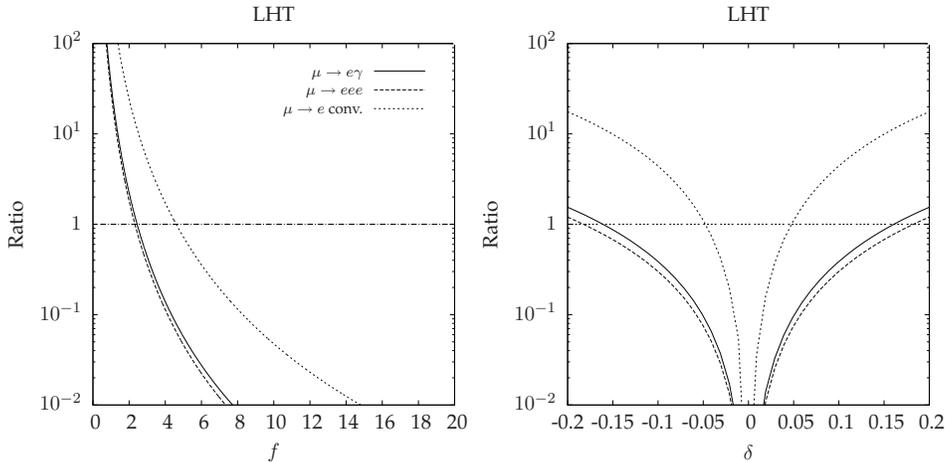}
\caption{Ratio of predictions to current experimental limits for the LHT model.}
\label{LHTplot}
\end{figure}

\begin{figure}[!htp]
\centering
\includegraphics[scale=0.71]{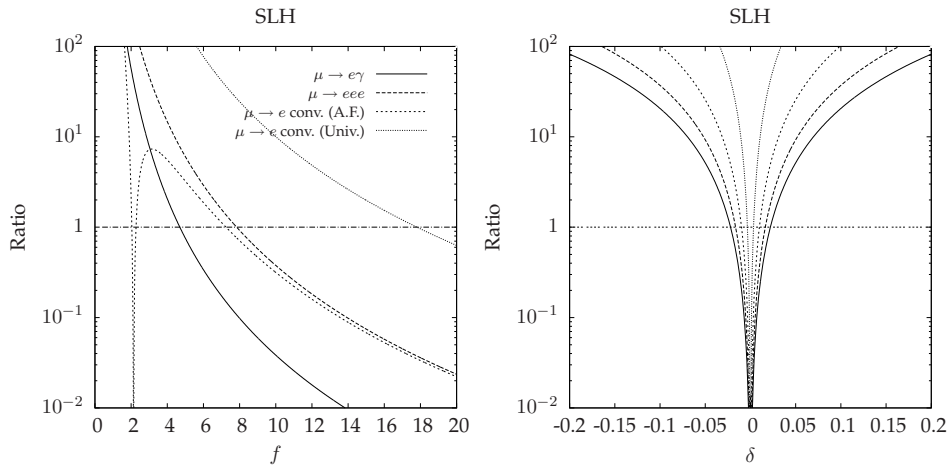}
\caption{Ratio of predictions to current experimental limits for the SLH model.}
\label{SLHplot}
\end{figure}

Figures \ref{LHTplot} and \ref{SLHplot} show the ratio of the Little Higgs predictions to the present experimental limits for the three lepton flavor violating processes under study.  A complete analysis can be found in \cite{delAguila:2008zu,newpaper}.  For the LHT model, there are complementary analyses by the Munich group (see \cite{Blanke:2009am} and references therein).

The form factors and amplitudes have an approximate behaviour given by:
\begin{equation}
 \mathcal{A} \propto \frac{v^2}{f^2} \sin 2\theta \; \delta
\end{equation}
The dependence on $\widetilde{y}$ is more involved and can not be expressed as simply.

In the LHT model we see that, in general, the process that gives the strongest restrictions on the parameter space is $\mu{\rm Ti} \to {\rm eTi}$.  We see that in order to get the prediction within the limit we have several options.  One can make the breaking scale $f$ somewhat heavier ($f\gtrsim 4.7$ TeV) or one can make either the mixing or mass splitting small, \ie, $\sin 2\theta \lesssim 0.05$ or $\delta \lesssim 5\%$.  These considerations make the model somewhat unnatural.  On the one hand, if you raise the scale you defeat the basic purpose of the model and, on the other, if you reduce the angles and splittings you are forcing a fine-tuning of the light and heavy mass matrices to within a few percent.  One would need some additional symmetry in the model to justify this.

In the SLH model the results are similar.  Note that here there are two possible \emph{embeddings} for the quark sector: the Universal embedding and the Anomaly Free embedding.  This produces different results in the case of $\mueN$ (see figure \ref{SLHplot}).  The heavier restrictions again come from the $\mu \to {\rm e}$ conversion process and they are much more extreme in the Universal embedding.  To fulfill the experimental constraints we need a heavy scale $f\gtrsim 8$ TeV or very small mixing angles or splittings, \ie, $\sin 2\theta \lesssim 0.01$ or $\delta \lesssim 1\%$.  The same considerations on naturalness that apply in the LHT model, apply in the SLH model.  However, the restrictions are more demanding in this case.

Certain very small regions of the parameter space seem to be allowed due to incidental cancellations of loop functions.  However, these regions are ruled out when all processes are taken into account.  In other words, these cancellations never happen in all three processes simultaneously.

\end{document}